\documentclass[12pt]{article}

\usepackage{epsf,epsfig}

\usepackage{a4wide}
\def\slash#1{#1 \hskip-0.45em /}
\def\Slash#1{#1 \hskip-0.59em /}

\def\beq{\begin{eqnarray}}
\def\eeq{\end{eqnarray}}

\def\be{\begin{equation}}
\def\ee{\end{equation}}

\def\np{n_+}
\def\nm{n_-}

\def\Wc{W_{c}}

\def\WZdag{W\! Z^\dagger}

\begin{document}
\thispagestyle{empty}

\begin{flushright}
  PITHA 02/17\\
  hep-ph/0211358\\
  November 21, 2002
\end{flushright}

\vspace{\baselineskip}

\begin{center}
\vspace{0.5\baselineskip}
\textbf{\Large 
Multipole-expanded soft-collinear effective theory\\[0.5em] 
with non-abelian gauge symmetry}\\
\vspace{3\baselineskip}
{\sc M. Beneke and Th.~Feldmann}\\
\vspace{2\baselineskip}
\textit{Institut f\"ur Theoretische Physik E, RWTH Aachen\\[0.1cm]
D -  52056 Aachen, Germany} \\
\vspace{3\baselineskip}

\vspace*{1cm}
\textbf{Abstract}\\
\vspace{1\baselineskip}
\parbox{0.9\textwidth}{
In position space the interaction terms of soft-collinear effective 
theory must be multipole-expanded to obtain interaction terms 
with homogeneous scaling behaviour. In this note we provide a
manifestly gauge-invariant formulation of the theory after this 
expansion in the presence of non-abelian gauge fields, extending 
our previous result. We give the effective Lagrangian (including the 
Yang-Mills Lagrangian for collinear and ultrasoft gluons) 
and heavy-to-light transition currents to
second order in the power expansion, paying particular attention to
the field redefinitions that lead to the gauge symmetries
of the effective Lagrangian. 
}
\end{center}

\newpage
\setcounter{page}{1}

%\tableofcontents

\newpage

%%%%%%%%%%%%%%%%%%%%%%%%%%%%%%%%%%%%%%%%%%%%%%%%%%%%%%%%%%%%%%%%%

%\section{Introduction}
\paragraph{} {\bf 1.} {\bf Introduction.} 
Soft-collinear effective theory (SCET) \cite{Bauer:2000yr} is a  
rapidly developing framework that allows us to simplify factorization
proofs for strong interaction processes which involve very energetic, 
nearly massless (``collinear'') particles. The effective theory is 
based on the observation that such processes contain at least the 
scales $Q$, $Q\lambda$ and $Q\lambda^2$, where $Q$ is the hard scale. 
Depending on the specific situation, either $Q\lambda$ or 
$Q\lambda^2$ is identified with the strong interaction scale, so that 
in any case $\lambda\ll 1$. 

The possibility to classify interactions in powers of $\lambda$ makes
SCET the appropriate framework to discuss power corrections to general
hard processes. The effective Lagrangian and certain ``currents'' 
were systematically studied to second order in $\lambda$  
in a position space formulation of SCET \cite{BCDF}, which in contrast to the
original hybrid momentum-position space representation does not make 
use of momentum ``label'' operators. However, in position space fields have to
be Taylor-expanded in directions  where they vary slowly, to obtain 
operators that scale with a definite power in $\lambda$. (We will
refer to this as ``homogeneous'' scaling behaviour as opposed to
operators that in addition to their leading $\lambda$ behaviour
contain a series of suppressed terms.) In \cite{BCDF} this
``multipole'' expansion has been performed, but a manifestly
gauge-invariant form of the effective Lagrangian and ``currents'' 
after multipole expansion has been derived only for abelian
gauge fields. In this note we complete
the construction of the homogeneous version of position-space SCET by
extending the previous result to non-abelian gauge fields. The
expansion will be performed to second order, but the procedure is
sufficiently general to make the construction of higher order terms a
straightforward exercise.

We recall \cite{Bauer:2000yr,BCDF} that the effective theory contains
a collinear quark field $\xi$, a collinear gluon field $A_c$, and 
ultrasoft quark and gluon fields, $q$ and $A_{\rm us}$, respectively. 
The components of collinear momentum scale as
$n_+ p \sim 1$, 
$p_\perp \sim \lambda$, 
$n_- p \sim \lambda^2$,
where $n_\pm^\mu$ are two light-like vectors, $n_+^2=n_-^2=0$ with
$\np\nm = 2$. In the position space formulation the arguments of
collinear fields scale as $\nm x\sim 1$, $x_\perp\sim 1/\lambda$, 
$\np x \sim 1/\lambda^2$. The components of ultrasoft momentum are all
of order $\lambda^2$ and hence ultrasoft fields vary only over 
large distances $x\sim 1/\lambda^2$. 
The components of collinear and ultrasoft gluon
fields scale as the components of the corresponding momentum, while 
$\xi\sim \lambda$ and $q\sim \lambda^3$. 
 
In \cite{BCDF} the effective theory was constructed in two
steps. First, fields satisfying the scalings specified above were
introduced and the corresponding Lagrangian was derived. This
Lagrangian was then multipole-expanded. The
effective Lagrangian before multipole expansion is invariant 
under a collinear and ultrasoft gauge symmetry (defined as gauge
symmetries where the gauge transformation $U(x)$ has the same
$x$ variations as collinear or ultrasoft fields) given by 
\begin{eqnarray}
&& 
\begin{array}{lll}
\mbox{collinear: }\quad & \displaystyle
   A_{c}\to U_{c} \, A_{c} \, U^{\dagger }_{c}
   + \frac{i}{g} \, U_{c}
   \left[D_{\rm us}, U^{\dagger }_{c}\right], \qquad & 
   \xi \to U_{c} \, \xi, \\[0.2cm] 
& A_{\rm us} \to A_{\rm us}, &
q \to q,\\[0.3cm] 
\mbox{ultrasoft: }\quad & 
   A_{c} \to  U_{\rm us} \, A_{c} \, U^{\dagger}_{\rm us}, & 
   \xi \to U_{\rm us} \, \xi, \\[0.1cm] 
&  \displaystyle
A_{\rm us} \to U_{\rm us}\, A_{\rm us} \, U^{\dagger}_{\rm us}
  + \frac{i}{g} \, U_{\rm us}
  \left[\partial, U^{\dagger}_{\rm us}\right], &
  q \to U_{\rm us} \, q.
\end{array}
\label{gaugetrafos}
\end{eqnarray}

Note while $A_c+A_{\rm us}$ transforms as the gauge field $A$ in full QCD
under both gauge symmetries (\ref{gaugetrafos}), $\xi+q$ does not
transform as the full QCD quark field $\psi$ under the collinear gauge
symmetry. The reason for this is that the effective Lagrangian is obtained 
after integrating out the two small components of the collinear quark
spinor and after applying the equation of motion for $\xi$. The
applications of the equation of motion are equivalent to a
redefinition of the $\xi$ field, after which the relation 
$\psi=F(\xi,q,A_c,A_{\rm us})$ is non-linear and non-local. 
On the other hand, the gauge field equation of motion is not used, so
that the relation between the effective and full gluon fields remains 
simple. In \cite{BCDF} this procedure based on a successive use of 
the equations of motion for $\xi$ 
has been used to determine the gauge-invariant
effective Lagrangian including second-order power-suppressed
interactions. Similarly, the relation between the QCD field and the
effective fields, $F$,  has been constructed to order $\lambda^2$. Then, 
if $\xi$ and $q$ transform according to (\ref{gaugetrafos}), 
$F(\xi,q,A_c,A_{\rm us})$ transforms as the full QCD field up to 
corrections of order $\lambda^3$ and higher. 

We may now extend the result of \cite{BCDF} to all orders in 
$\lambda$ by making a further field redefinition (affecting only 
terms in the action smaller than $\lambda^2$) such that 
$F(\xi,q,A_c,A_{\rm us})$ 
transforms as the full QCD field exactly. Such a redefinition
can always be made. In the approach of \cite{BCDF} this corresponds 
to further applications of the equation of motion for $\xi$, although
the explicit construction can become rather difficult. After this 
redefinition the QCD fields are related to the effective fields by  
\begin{eqnarray}
\label{adef} A &=& A_c+A_{\rm us},\\
\label{psidef} \psi &=& \xi+\WZdag q-\frac{1}{i n_+ D}\,\frac{\slash n_+}{2} 
\left(i\Slash D_\perp \xi + \left[\left[i\Slash D_\perp\WZdag
\right]\right] q\right)
\end{eqnarray}
exactly. Here $W$ and $Z$ are Wilson lines, 
\begin{equation}
\WZdag(x) \equiv  P\exp\left(ig\int_{-\infty}^0 \!ds \,n_+ A(x+s
    n_+)\right)  \bar P\exp\left(-ig\int_{-\infty}^0 \!ds \,n_+ A_{\rm
      us}(x+s n_+)\right),
\end{equation}
where $P$ denotes path-ordering 
and $\bar P$  reverse path-ordering. We also used the 
``double-bracket'', defined by 
\begin{equation}
\left[\left[f(D) A\right]\right] \equiv f(D) A-A f(D_{\rm us}),
\qquad  
\left[\left[A f(D)\right]\right] \equiv A f(D)-f(D_{\rm us}) A, 
\end{equation}
and $i D = i\partial +g (A_c+A_{\rm us})$, 
$i D_{\rm us} = i\partial +g A_{\rm us}$. 
Then, if the effective fields 
transform according to (\ref{gaugetrafos}), $A$ and $\psi$ transform
as the gauge and quark field in full QCD under collinear and ultrasoft
gauge transformations. 

The SCET Lagrangian before multipole expansion given in
\cite{BCDF} can now be derived very easily and extended to all orders 
in $\lambda$ by inserting the 
field redefinitions (\ref{adef},\ref{psidef}) into the QCD
Lagrangian. We obtain 
\begin{eqnarray}
\label{collfinal}
{\cal L} &=&  
\bar\xi\, \left(i n_- D + i \Slash D_\perp \frac{1}{i n_+ D}\,
         i\Slash D_\perp \right)  \frac{\slash n_+}{2} \xi 
+ \bar{q}\, i \Slash{D}_{\rm us}\, q 
%\nonumber\\[0.2cm]
%&& +\,
+ \bar \xi \left[\left[i \Slash D_\perp \WZdag\right]\right] q + 
\bar q \left[\left[Z W^\dagger  i \Slash D_{\perp}\right]\right] \xi 
\nonumber\\[0.2cm]
&& +\,\bar \xi\,\frac{\slash n_+}{2} \left[\left[i\nm D \,\WZdag 
\right]\right] q 
+ \bar\xi \,\frac{\slash n_+}{2}\,i \Slash D_\perp\, \frac{1}{in_+ D}\,
 \left[\left[i \Slash D_\perp \WZdag \right]\right] q 
\nonumber\\[0.2cm]
&& +\,  \bar q \left[\left[Z W^\dagger \,i\nm D \right]\right] 
\frac{\slash n_+}{2} \xi 
+\bar q \,  \left[\left[Z W^\dagger i \Slash D_\perp \right]\right] 
\, \frac{1}{in_+ D}\, i \Slash D_\perp
\frac{\slash n_+}{2}\xi 
\nonumber\\[0.2cm]
&& -\,\bar q  
\left[\left[Z W^\dagger i \Slash D_\perp\right]\right]\frac{\slash
  n_+}{2}\, \frac{1}{in_+ D}\, 
\left[\left[i \Slash D_\perp \WZdag \right]\right] q.
\end{eqnarray}
To derive this one must use ``momentum conservation'', which means
that terms with a single collinear field (multiplied by ultrasoft
fields) can be dropped, since they do not contribute to the action. 
In \cite{BCDF} momentum conservation has been used to drop terms 
of the form $\bar \xi i \Slash D_{\perp\rm us} q$. This is
inconvenient, because single collinear fields can be transformed into
composite collinear fields by collinear gauge
transformations. Manifest gauge invariance is then only restored after
further applications of the equations of motion as can be seen in
\cite{BCDF}. We can render this procedure manifestly gauge invariant
by choosing a reference gauge in which momentum conservation is
applied. Then, in any gauge, we first go to the reference gauge by
applying the gauge transformation $U_c$, use
momentum conservation in this gauge, and then transform back to the
original gauge with $U_c^\dagger$. We choose collinear light-cone
gauge as our reference gauge, in which case $U_c=ZW^\dagger$. Then 
momentum conservation allows us to add or drop terms such as 
\begin{equation}
\bar \xi \,\WZdag i \Slash D_{\perp\rm us} q,\quad
\bar q Z W^\dagger 
\left[\left[i \Slash D_\perp \WZdag\right]\right] q,
\end{equation}
because $\WZdag=1$ in collinear light-cone gauge $\np A_c=0$. 
In particular $\bar q  \,Z W^\dagger i \Slash D\,\WZdag q$ is equivalent 
to $\bar{q}\, i \Slash{D}_{\rm us}\, q$.

The last term in (\ref{collfinal}) and an additional 
term $\bar\xi \,\WZdag i \Slash D_{\rm us} q \,+ \,\mbox{h.c.}$,
contained in the double brackets, have not been given in \cite{BCDF}, 
because they are $\lambda^4$ corrections. With the addition of these  
terms the effective Lagrangian is exact to all orders in
$\lambda$, and this may be considered an advantage of this formalism. 
(The pure gluon Lagrangian is 
at this stage the same as in full QCD.) However, the
individual terms in the Lagrangian do not have a
simple $\lambda$ scaling due to the presence of $A_c$ and 
$A_{\rm us}$ in the covariant derivative and in $W$, 
although their largest component is easily
determined. The inhomogeneity is inevitable at this stage, because the
gauge transformations are not homogeneous. The inhomogeneity arises 
form the presence of $A_{\rm us}$ in the collinear transformation of 
$A_c$, and from the multiplication of collinear fields with the
ultrasoft function $U_{\rm us}(x)$. However, the exact Lagrangian
(\ref{collfinal}) serves as a starting point for the multipole expansion.

\paragraph{} {\bf 2.} {\bf Homogeneous gauge transformations.}
The effective theory should be constructed such that every term has a
simple (homogeneous) scaling behaviour. In addition to expanding 
quantities such as $\WZdag$ and $(i\np D)^{-1}$ in (\ref{collfinal})
one must account for the fact that momentum is not conserved at
collinear-ultrasoft interaction vertices. To be specific, when an 
incoming  collinear line with momentum $p$ absorbs an ultrasoft momentum
$k$, the outgoing collinear line has momentum $p+\frac{1}{2}(\nm k)
\np$. The components of $k$ small relative to those of  
$p$ are neglected in the
propagator; the corresponding terms in the expansion of the full
propagator are part of interaction vertices. In position space this
corresponds to the Taylor-expansion of ultrasoft fields around 
$x_- \equiv \frac{1}{2} (n_+ x)\,n_-$, whenever they multiply
collinear fields, since in such a product the $x$ variations are
dominated by the variations of collinear fields. (The procedure is
analogous to the familiar multipole expansion in atomic physics, and 
in non-relativistic effective field theory \cite{nrqcd}, where the role of 
light-front time $x_-$ is taken by real time $t$.) We therefore perform
the ``light-front multipole expansion'' \cite{BCDF} 
\begin{eqnarray}
\phi_{\rm us}(x) &=& 
  \phi_{\rm us}(x_-) + \Big[x_{\perp} \partial
\phi_{\rm us}\Big](x_-)
\nonumber \\
&& + \,\frac{1}{2} \,n_- x \Big[n_+ \partial \,\phi_{\rm us}\Big](x_-)
 + \frac{1}{2} \Big[x_{\mu \perp} x_{\nu \perp}
      \partial^\mu\partial^\nu \phi_{\rm us} \Big](x_-)
 + {\cal O}(\lambda^3 \phi_{\rm us}).
  \label{taylor}
\end{eqnarray}
of all ultrasoft fields, where $x_- = \frac{1}{2} (n_+ x)\,
n_-$. The expanded effective Lagrangian is homogeneous in $\lambda$,
but it is obviously no longer invariant term by term under 
the gauge transformations 
(\ref{gaugetrafos}), since they mix different orders in $\lambda$.

For abelian gauge fields it was shown 
\cite{BCDF} that the effective Lagrangian 
does assume a gauge invariant form after several applications of the 
equation of motion for $\xi$, 
after which the collinear quark
field $\xi$ transforms with $U_{\rm us}(x_-)$ under ultrasoft 
transformations $U_{\rm us}(x)$. 
The applications of the equation of
motion for $\xi$ are equivalent to the field redefinition
\begin{equation}
\xi(x) = \exp\left(ig\int_C dy_\mu A^\mu_{\rm us}(y)\right)\hat{\xi}(x),
\label{Rabelian}
\end{equation}
where $C$ denotes a straight path from $x_-$ to $x$. The new field 
$\hat{\xi}(x)$ has the homogeneous transformation  
$\hat{\xi}(x)\to U_{\rm us}(x_-)\hat{\xi}(x)$, guaranteeing the 
term-by-term gauge invariance of the multipole-expanded Lagrangian. 
The abelian case is simple, because collinear
gauge transformations {\em are} homogeneous in $\lambda$ for abelian 
gauge fields and $A_c$ does not transform under ultrasoft gauge 
transformations. Hence, for abelian fields the only inhomogeneity in 
(\ref{gaugetrafos}) comes from $U_{\rm us}$ when it multiplies the 
collinear field $\xi$. 

{}From this it is clear that in the non-abelian case we must find 
new collinear fields $\hat{\xi}$ and $\hat{A}_c$, such that the
Lagrangian expressed in terms of the new field variables is invariant
under the homogenized version of the gauge symmetries, given by 
\begin{eqnarray}
\begin{array}{ll}
\mbox{\hspace*{-1cm} collinear: } & \\ 
   \displaystyle
   \np \hat{A}_{c}\to U_{c} \, \np \hat{A}_{c} \, U^{\dagger }_{c}
   + \frac{i}{g} \, U_{c}
   \left[\np\partial, U^{\dagger }_{c}\right], \qquad & 
   \hat{\xi} \to U_{c} \, \hat{\xi},\\[0.2cm] 
  \displaystyle
\hat{A}_{\perp c}\to U_{c} \, \hat{A}_{\perp c} \, U^{\dagger }_{c}
   + \frac{i}{g} \, U_{c}
   \left[\partial_{\perp}, U^{\dagger }_{c}\right], \qquad & 
    \\[0.2cm] 
 \displaystyle
\nm \hat{A}_{c}\to U_{c} \, \nm \hat{A}_{c} \, U^{\dagger }_{c}
   + \frac{i}{g} \, U_{c}
   \left[\nm D_{\rm us}(x_-), U^{\dagger }_{c}\right], \qquad & 
  \\[0.4cm] 
 A_{\rm us} \to A_{\rm us}, &
q \to q,\\[0.3cm] 
\mbox{\hspace*{-1cm} ultrasoft: } & \\[0.2cm]
   \hat{A}_{c} \to  U_{\rm us}(x_-) \, 
   \hat{A}_{c} \, U^{\dagger}_{\rm us}(x_-), & 
   \hat{\xi} \to U_{\rm us}(x_-) \, \hat{\xi}, \\[0.2cm] 
  \displaystyle
  A_{\rm us} \to U_{\rm us}\, A_{\rm us} \, U^{\dagger}_{\rm us}
  + \frac{i}{g} \, U_{\rm us}
  \left[\partial, U^{\dagger}_{\rm us}\right], &
  q \to U_{\rm us} \, q.
\end{array}
\label{newgaugetrafos}
\end{eqnarray}
It is easily checked that every term now has the same scaling in
$\lambda$ as was required. Since the transformations of ultrasoft
fields are unaltered, no redefinition of these fields is needed. 
In (\ref{newgaugetrafos}) fields and gauge
transformations without argument are taken at $x$ as in 
(\ref{gaugetrafos}), while other arguments are given explicitly.  

Note that the collinear Wilson line 
\begin{equation}
\label{wline/def}
  W_c(x) \equiv P\exp\left(ig\int_{-\infty}^0 \!ds \,n_+ \hat{A}_c(x+s
    n_+)\right)
\end{equation} 
has the simple transformations 
\begin{equation}
W_c\to U_c W_c,\qquad W_c\to  U_{\rm us}(x_-) W_c U^{\dagger}_{\rm us}(x_-), 
\end{equation}
because the arguments of collinear fields in the path-ordered product 
correspond to the {\em same} $(x+s\np)_-=x_-$. The other objects with
simple transformation properties under (\ref{newgaugetrafos}) are 
$\hat{\xi}$, $q$, $F^{\mu\nu}_{\rm us}$,
$i\np \hat{D}_c$, $i \hat{D}_{\perp c}$, $i\nm \hat{D}$ 
(but not $i\nm \hat{D}_c$) and 
$i D_{\rm us}^\mu$. (The ``hat'' indicates that the covariant
derivative contains $\hat{A}_c$, not $A_c$.) 
The multipole-expanded Lagrangian will be composed
of these objects.

\paragraph{} {\bf 3.} {\bf Field redefinitions.} 
To find the field variables that lead to the new gauge symmetries,
we first fix the collinear gauge symmetry by choosing light-cone gauge
for $A_c$ and $\hat{A}_c$, such that $\np A_c=\np\hat{A}_c=0$. We then
define 
\begin{equation}
\label{LCdef}
\xi = R \,\hat{\xi},\qquad A_c = R\hat{A}_c R^\dagger,
\end{equation}
where now
\begin{equation}
\label{Rdef}
R(x) = P\exp\left(ig\int_C dy_\mu A^\mu_{\rm us}(y)\right)
\end{equation}
with $C$ a straight path from $x_-$ to $x$. Since the distance from 
$x_-$ to $x$ is at most of order $1/\lambda$, whereas $A_{\rm us}$
varies only over distances of order $1/\lambda^2$, we can expand $R$ in 
$\lambda$ using 
\begin{eqnarray}
&& \int_C dy_\mu A^\mu_{\rm us}(y) = \int_0^1 ds\,(x-x_-)_\mu 
A^\mu_{\rm us}(x_-+s(x-x_-))\nonumber\\
&&\hspace*{1.5cm} = x_{\perp\mu} A^\mu_{\rm us}(x_-) + 
\frac{1}{2} \nm x\,\np A_{\rm us}(x_-) +
\frac{1}{2} x_{\perp\mu} x_{\perp\nu} \left[\partial^\nu
  \!A^\mu_{\rm us}
\right](x_-)
+\ldots,
\end{eqnarray}
where after the second equality the first term is of order $\lambda$ and the
other two of order $\lambda^2$, and 
all fields are evaluated at $x_-$ ({\em after}
derivatives are taken). It is straightforward to see that the new
fields $\hat{\xi}$, $\hat{A}_c$ have the required transformations 
(\ref{newgaugetrafos}) under the ultrasoft gauge symmetry. 

We now continue to assume collinear light-cone gauge for $A_c$, 
but we restore collinear gauge invariance for the new fields. To this
end, we note that if $\hat{A}_c$ is not in light-cone gauge, a gauge
transformation $U_c=W_c^\dagger$ will transform it to this
gauge. Hence, we should replace in (\ref{LCdef}) $\hat{\xi}$ by 
$W_c^\dagger \hat{\xi}$ and $g\hat{A}_{\perp c}$ by 
$W_c^\dagger [i\hat{D}_c W_c]$ etc.\ (since according to
(\ref{newgaugetrafos}) these {\em are\/} the quark and gluons fields in 
collinear light-cone gauge). We then find the following collinear field
redefinitions: 
\begin{eqnarray}
\label{redefcoll}
\xi &=& R W_c^\dagger \hat{\xi}, \nonumber\\
g A_{\perp c} &=& % R W_c^\dagger [iD_{\perp c} W_c] R^\dagger = 
 R\left(W_c^\dagger i\hat{D}_{\perp c} W_c - i\partial_\perp\right) R^\dagger,
 \\
g \nm A_{c} &=& 
%R \left(W_c^\dagger [i\nm D W_c] -  g \nm A_{\rm us} \right)R^\dagger = 
 R\left(W_c^\dagger i\nm \hat{D} W_c - i\nm D_{\rm us}(x_-)\right) R^\dagger.
\nonumber
\end{eqnarray}
Recall that the fields without hats on the left-hand side are still in
light-cone gauge. The
quantities on the right-hand side are expressed entirely in terms of 
the new collinear gluon field $\hat{A}_c$. It
is straightforward to verify that the new fields have the
required collinear and ultrasoft transformations
(\ref{newgaugetrafos}). In particular, when the fields on the
right-hand side transform according to the collinear gauge symmetry 
(\ref{newgaugetrafos}),
the expressions in (\ref{redefcoll}) remain invariant as they should,
because the collinear gauge symmetry is fixed for the fields on the 
left-hand side.

\paragraph{} {\bf 4.} {\bf The multipole-expanded quark Lagrangian.}  
For the remainder of this paper we use the
following notation: 
we drop the ``hat'' on the new fields, since the multipole-expanded
SCET Lagrangian contains only these fields. Collinear fields 
without argument will be understood to be evaluated at $x$, but
ultrasoft fields without arguments are always evaluated at $x_-= 
\frac{1}{2} (n_+ x)\,n_-$. Furthermore, derivatives on ultrasoft 
fields operate on the field before setting $x=x_-$; derivatives 
enclosed in square brackets operate only
inside the bracket.

The field redefinitions (\ref{redefcoll}) are inserted into
(\ref{collfinal}) taken in collinear 
light-cone gauge (where $\WZdag=1$) and the resulting expression is
expanded in $\lambda$. To see the sort of terms that arise, we
consider the collinear Lagrangian given by the first term in
(\ref{collfinal}), which takes the form 
\begin{eqnarray}
{\cal L} &=&  
\bar{\xi}\,i n_- D \frac{\slash n_+}{2} \xi + 
   \bar{\xi}\, W_c\left(R^\dagger i\nm D_{\rm us}(x)R-i\nm
   D_{\rm us}\right)W_c^\dagger \frac{\slash n_+}{2} \xi
\nonumber\\ 
&&+\,\bar{\xi}\,\left(i\Slash D_{\perp c}+
W_c \left(R^\dagger i\Slash D_{\perp \rm us}(x) R-i\Slash\partial_\perp
\right)W_c^\dagger\right) W_c R^\dagger \frac{1}{i n_+ D_{\rm us}(x)}
R W_c^\dagger 
\nonumber\\
&&\hspace*{0.4cm} \left(i\Slash D_{\perp c}+
W_c \left(R^\dagger i\Slash D_{\perp \rm us }(x) R-i\Slash\partial_\perp
\right)W_c^\dagger\right) \frac{\slash n_+}{2} \xi
\end{eqnarray}
when expressed in the new field variables. (Note that with our
conventions $i n_- D$ contains the collinear gauge field at $x$ and 
the ultrasoft gauge field at $x_-$.) From this and similar 
manipulations of the terms in the
Lagrangian with the ultrasoft quark field, we see that we need the
expansion of $(R^\dagger i\nm D_{\rm us}(x)R-i\nm D_{\rm us})$, 
$(R^\dagger i\Slash D_{\perp \rm us}(x) R-i\Slash\partial_\perp)$, 
$R^\dagger (i n_+ D_{\rm us}(x))^{-1} R$, 
$R^\dagger q(x)$ and $i\Slash\partial_\perp R^\dagger q(x)$ in 
$\lambda$. 

The expansion is constructed most easily by choosing the special 
gauge 
\begin{equation}
(x-x_-)_\mu A^\mu_{\rm us}(x) = 
\left(x_{\perp\mu}+\frac{1}{2} \,(\nm x) 
\,{\np}_\mu\right) A^\mu_{\rm us}(x) =0,
\end{equation}
which by analogy with Fock-Schwinger or fixed-point gauge we will
refer to as fixed-line gauge (as $x_-$ depends on $x$ through $\np
x$). If $A^\mu_{\rm
  us}(x)$ does not satisfy the gauge condition, we can always choose 
$U_{\rm us}(x)=R^\dagger$ to transform the field to fixed-line
gauge. In fixed-line gauge the gauge field can be represented in terms
of the field strength tensor by the relations
\begin{eqnarray}
\label{flfields}
&& \nm A_{\rm us}(x) - \nm A_{\rm us} = 
\int_0^1 ds\,(x-x_-)^\mu n_-^\nu F_{\mu\nu}^{\rm us}(y(s)),
\nonumber\\
&& \np A_{\rm us}(x)  = 
\int_0^1 ds\,s\,(x-x_-)^\mu n_+^\nu F_{\mu\nu}^{\rm us}(y(s)),
\\
&& A_{\rm us \, \nu_\perp}(x)  = 
\int_0^1 ds\,s\,(x-x_-)^\mu  F_{\mu\nu_\perp}^{\rm us}(y(s)),
\nonumber
\end{eqnarray}
with $y(s)= x_- + s (x-x_-)$ parameterizing a straight path from $x_-$
to $x$. (The subscript ``$\perp$'' on an index means that a transverse
projection is done in this index.) From this we also deduce 
\begin{equation}
\label{gaugezeros}
\np A_{\rm us}(x_-)=A_{\perp\rm us}(x_-)=0. 
\end{equation}
in fixed-line gauge. 
The relation with the expressions above is now established, since 
we can use the ultrasoft gauge transformation $U_{\rm us}^\dagger(x) 
=R$ to return to the general gauge. This converts 
$F^{\mu\nu}_{\rm us}(y(s))$ to $ R^\dagger(y(s)) \, 
F_{\mu\nu}^{\rm us}(y(s)) \, R(y(s))$ and, for instance, 
$g\Slash A_{\perp \rm us}(x)$ to 
$(R^\dagger i\Slash D_{\perp \rm us}(x) R-i\Slash\partial_\perp)$, which
we needed. 
We therefore obtain
\begin{eqnarray}
&& R^\dagger i\nm D_{\rm us}(x)R-i\nm D_{\rm us} 
= \int_0^1 ds\,(x-x_-)^\mu n_-^\nu \, R^\dagger(y(s)) \, 
gF_{\mu\nu}^{\rm us}(y(s)) \, R(y(s))
\nonumber\\
&&\hspace*{0.4cm}=
\sum_{n=0}^\infty \frac{1}{(n+1)!} \, (x-x_-)^\mu 
(x-x_-)_{\rho_1}\ldots (x-x_-)_{\rho_n}
\big[D_{\rm us}^{\rho_1}, \big[D_{\rm us}^{\rho_2},\ldots
\big[D_{\rm us}^{\rho_n}, n_-^\nu gF_{\mu\nu}^{\rm us}\big]\ldots\big]\big]
\nonumber\\
&&\hspace*{0.4cm}=
x_{\perp}^{\mu}n_-^\nu gF_{\mu\nu}^{\rm us} + 
\frac{1}{2}\,\nm x \,n_+^\mu n_-^\nu gF_{\mu\nu}^{\rm us} + 
\frac{1}{2} \,x_{\perp}^{\mu}x_{\perp\rho} n_-^\nu \big[D_{\rm us}^\rho,
gF_{\mu\nu}^{\rm us}\big] + {\cal O}(\lambda^5),
\\[0.4em]
&& R^\dagger i\Slash D_{\perp \rm us}(x)R-i\Slash\partial_\perp
= \int_0^1 ds\,s\,(x-x_-)^\mu \gamma_{\perp}^{\nu} \, R^\dagger(y(s)) \,
g F_{\mu\nu}^{\rm us}(y(s)) \, R(y(s))
\nonumber\\
&&\hspace*{0.4cm}=
\sum_{n=0}^\infty \frac{1}{n!(n+2)} \,(x-x_-)^\mu 
(x-x_-)_{\rho_1}\ldots (x-x_-)_{\rho_n}
\big[D_{\rm us}^{\rho_1}, \big[D_{\rm us}^{\rho_2},\ldots
\big[D_{\rm us}^{\rho_n}, \gamma_{\perp}^{\nu}
g F_{\mu\nu}^{\rm us}\big]\ldots\big]\big]
\nonumber\\
&&\hspace*{0.4cm}=
\frac{1}{2}\,x_{\perp}^{\mu}\gamma_{\perp}^{\nu}
g F_{\mu\nu}^{\rm us} + {\cal O}(\lambda^4),
\\[0.4em]
&&R^\dagger \frac{1}{i n_+ D_{\rm us}(x)} R
= 
 \frac{1}{i n_+ \partial} - 
 \frac{1}{i n_+ \partial} \, \frac{1}{2} \,x_{\perp}^{\mu} n_+^\nu 
  \, gF_{\mu\nu}^{\rm us} \, \frac{1}{i n_+ \partial} 
+ {\cal O}(\lambda^4),
\\[0.4em]
&& R^\dagger q(x) = \sum_{n=0}^\infty \frac{1}{n!} \,
(x-x_-)_{\rho_1}\ldots (x-x_-)_{\rho_n}
D_{\rm us}^{\rho_1}\ldots D_{\rm us}^{\rho_n} q
\nonumber\\
&&\hspace*{0.4cm}=
q+x_{\perp\mu} D_{\rm us}^\mu q + {\cal O}(\lambda^2 q),
\\[0.4em]
&& i\Slash\partial_\perp R^\dagger q(x) = 
\sum_{n=0}^\infty \frac{1}{n!} \,
(x-x_-)_{\rho_1}\ldots (x-x_-)_{\rho_n}
i\Slash D_{\perp \rm us} D_{\rm us}^{\rho_1}\ldots D_{\rm us}^{\rho_n} q,
\end{eqnarray}
The expansion in terms of covariant derivatives is obtained most
directly from the Taylor-expansion of the fields and integrals in
fixed-line gauge, since
(\ref{gaugezeros}) allows us to convert the ordinary derivatives 
$\partial_\perp$ and $\np \partial$ to
covariant ones at $x_-$ in this gauge. Only after this expansion one 
returns to the general gauge, which becomes trivial since $R(x_-)=1$. 
After this expansion every single term has a
homogeneous scaling behaviour in $\lambda$.

With these results it is easy to write down the multipole-expanded
SCET Lagrangian to any order in $\lambda$. 
To order $\lambda^2$ the result takes the form 
\be
\label{fivet}
{\cal L} = \bar{\xi} \left(i n_- D + i \Slash{D}_{\perp c}
 \frac{1}{i n_+ D_{c}}\, i\Slash{D}_{\perp c} \right)
 \frac{\slash{n}_+}{2} \, \xi + 
 \bar{q}(x)\, i \Slash{D}_{\rm us}(x) q(x) 
 + {\cal L}_\xi^{(1)} + {\cal L}_\xi^{(2)} + 
 {\cal L}_{\xi q}^{(1)} + {\cal L}_{\xi q}^{(2)},
\ee
where the power-suppressed interaction terms are given by
\begin{eqnarray}
\label{corr1}
{\cal L}^{(1)}_{\xi} &=&
  \bar{\xi} \left( x_\perp^\mu n_-^\nu \, W_c \,gF_{\mu\nu}^{\rm
      us}W_c^\dagger \right) \frac{\slash n_+}{2} \, \xi,
\\[0.2cm]
{\cal L}^{(2)}_{\xi} &=&
  \frac{1}{2} \, \bar \xi \left(
  (n_-x) \, n_+^\mu n_-^\nu \, W_c \,gF_{\mu\nu}^{\rm us}  W_c^\dagger
  + x_\perp^\mu x_{\perp\rho} n_-^\nu W_c \big[D^\rho_{\rm us}, 
   g F_{\mu\nu}^{\rm us}\big] W_c^\dagger  \right) 
   \frac{\slash  n_+}{2} \, \xi
\\
&& + \,\frac{1}{2} \, \bar \xi \left(
    i \Slash D_{\perp c}  \,
 \frac{1}{i n_+ D_{c}} \, x_\perp^\mu \gamma_\perp^\nu \,
 W_c \,g F_{\mu\nu}^{\rm us}W_c^\dagger  +   x_\perp^\mu \gamma_\perp^\nu \,
 W_c \,g F_{\mu\nu}^{\rm us}W_c^\dagger  \,
 \frac{1}{i n_+ D_{c}} \, i \Slash D_{\perp c}
 \right) \frac{\slash n_+}{2}
\, \xi,
\nonumber\\[0.2cm]
{\cal L}^{(1)}_{\xi q} &=& 
    \bar q \,W_{c}^\dagger i\Slash{D}_{\perp c} \,\xi - 
    \bar{\xi} \,i\overleftarrow{\Slash D}_{\perp c} W_{c}\, q , 
\\[0.2cm]
{\cal L}^{(2)}_{\xi q} &=&  
\bar{q} \,\Wc^\dagger \left( i n_- D
   + i \Slash{D}_{\perp c}\,
     \left(i n_+ D_{c}\right)^{-1} i \Slash{D}_{\perp c}\right)
     \frac{\slash{n}_+}{2} \,\xi
+ \, \bar q \,x_{\perp\mu}
  \overleftarrow{D}^\mu_{\rm us} W_{c}^\dagger \,i \Slash D_{\perp c} \xi 
\\
&& - \, \bar{\xi}\,\frac{\slash{n}_+}{2} \left( i n_- \overleftarrow{D}
   + i \overleftarrow{\Slash D}_{\perp c}\,
     \left(i n_+ \overleftarrow{D}_{c}\right)^{-1}
    i \overleftarrow{\Slash D}_{\perp c}\right) \Wc\, q
- \, \bar \xi \, i \overleftarrow{\Slash D}_{\perp c} W_{c} \,x_{\perp\mu}
  D^\mu_{\rm us} q.
\label{corr2}
\end{eqnarray}
For abelian gauge fields these 
expressions coincide with those given
in \cite{BCDF}. (The final result (\ref{corr1}-\ref{corr2}) 
for the non-abelian case has already been presented in 
\cite{Feldmann:2002cm}.) In this form every term in the Lagrangian
scales as a single power in $\lambda$, which can be determined from the
scaling rules for fields, coordinates and derivatives. We note that
this Lagrangian is exact, i.e. its coefficients are not modified by
radiative corrections, neither do radiative corrections induce 
new operators \cite{BCDF}.

\paragraph{} {\bf 5.} {\bf The Yang-Mills Lagrangian.} 
It remains to perform the $\lambda$ expansion of the pure Yang-Mills 
Lagrangian. Recall that in the first version of position space SCET,
which is invariant under the gauge symmetry (\ref{gaugetrafos}), the 
Yang-Mills part of the Lagrangian is the same as in QCD with 
$A$ replaced by $A_c+A_{\rm us}$. Before the
field redefinition (\ref{redefcoll}) we can rearrange the Yang-Mills
Lagrangian as
\begin{equation}
\label{YM1}
{\cal L}_{\rm YM} = -\frac{1}{2}\,\mbox{tr}\left(G_c^{\mu\nu}
  G^c_{\mu\nu}\right) - \mbox{tr}\left(G_c^{\mu\nu}
  F^{\rm us}_{\mu\nu}(x)\right) - 
  \frac{1}{2}\,\mbox{tr}\left(F_{\rm us}^{\mu\nu}(x)
  F^{\rm us}_{\mu\nu}(x)\right)
\end{equation}
with the definition
\begin{equation}
G_c^{\mu\nu} = [D_{\rm us}^\mu(x),A_c^\nu] - 
[D_{\rm us}^\nu(x),A_c^\mu] -i g\,[A_c^\mu,A_c^\nu].
\end{equation}
The first two terms of (\ref{YM1}) are products of collinear and
ultrasoft fields which must be multipole-expanded. The third term is
the ultrasoft Yang-Mills Lagrangian, which contributes a leading power
term to the action. 

As before we go to light-cone gauge for the collinear fields, insert
the field redefinitions (\ref{redefcoll}) and expand the result in
$\lambda$ choosing ultrasoft fixed-line gauge in an
intermediate step. Including terms of order $\lambda^2$ the result is 
\begin{equation}
{\cal L}_{\rm YM} = 
 -\frac{1}{2}\,\mbox{tr}\left(F_c^{\mu\nu}
  F^c_{\mu\nu}\right) - 
  \frac{1}{2}\,\mbox{tr}\left(F_{\rm us}^{\mu\nu}(x)
  F^{\rm us}_{\mu\nu}(x)\right) 
 +{\cal L}^{(1)}_{\rm YM} +{\cal L}^{(2)}_{\rm YM},
\end{equation}
where we define the collinear field strength tensor $F^c_{\mu\nu}$ 
through its components 
\begin{eqnarray}
&&g n_{+\mu} n_{-\nu} F_c^{\mu\nu} \equiv [\np D_c,i\nm D],\qquad
g F_c^{\mu_\perp\nu_\perp} \equiv [D_{c}^{\mu_\perp},iD_{c}^{\nu_\perp}],
\nonumber\\ 
&&g n_{+\mu} F_c^{\mu\nu_\perp} \equiv [\np D_c,iD_{c}^{\nu_\perp}],\qquad 
\hspace*{0.3cm}
g n_{-\mu} F_c^{\mu\nu_\perp} \equiv [\nm D,iD_{c}^{\nu_\perp}]. 
\end{eqnarray}
This definition almost coincides with the standard one except that it
contains $\nm D$ rather than $\nm D_c$, which is related to the
presence of $A_{\rm us}$ in the collinear transformation of 
$\nm \hat{A}_c$ in (\ref{newgaugetrafos}). The first and
second order 
power-suppressed gluon self-interactions are given by
\begin{eqnarray}
{\cal L}^{(1)}_{\rm YM} &=& 
\mbox{tr}\left(n_+^\mu F^c_{\mu\nu_\perp} \! W_c \,i\Big[x_\perp^\rho 
n_-^\sigma F^{\rm us}_{\rho\sigma},W_c^\dagger [iD_c^{\nu_\perp} W_c]
\Big] W_c^\dagger\right)
- \mbox{tr}\left(n_{+\mu} F_c^{\mu\nu_\perp} W_c n_-^\rho F^{\rm us}_{\rho
    \nu_\perp} W_c^\dagger \right)\!,
\cr &&
\\[0.0cm]
{\cal L}^{(2)}_{\rm YM} &=& 
\frac{1}{2} \,\mbox{tr}\left(n_+^\mu F^c_{\mu\nu_\perp} W_c 
\,i\Big[\nm x\,n_+^\rho n_-^\sigma F^{\rm us}_{\rho\sigma} + 
x_\perp^\rho x_{\perp\omega} n_-^\sigma 
\big[D_{\rm us}^\omega,F^{\rm us}_{\rho\sigma}\big],
W_c^\dagger [iD_c^{\nu_\perp} W_c]
\Big] W_c^\dagger\right)
\nonumber\\[0.1cm]
&&-\,\frac{1}{2} \,\mbox{tr}\left(n_{+\mu} F_c^{\mu\nu_\perp} W_c 
\,i\Big[x_\perp^\rho n_-^\sigma F^{\rm us}_{\rho\nu_\perp},
W_c^\dagger i\nm D W_c - i\nm D_{\rm us}
\Big] W_c^\dagger\right)
\nonumber\\[0.1cm]
&&+ \,\mbox{tr}\left(F_c^{\mu_\perp\nu_\perp} W_c 
\, i\Big[x_\perp^\rho  F^{\rm us}_{\rho\mu_\perp},
W_c^\dagger [iD_{c \, \nu_\perp} W_c]\Big] W_c^\dagger\right)
\nonumber\\[0.1cm]
&&+\,\frac{1}{2} \,\mbox{tr}\left(n_+^\mu n_-^\nu F^c_{\mu\nu}
W_c n_+^\rho n_-^\sigma F^{\rm us}_{\rho\sigma}W_c^\dagger \right) - 
\mbox{tr}\left(F_c^{\mu_\perp\nu_\perp} W_c F^{\rm us}_{\mu_\perp\nu_\perp}
W_c^\dagger\right)
\nonumber\\[0.1cm]
&&-\,\mbox{tr}\left(n_{+\mu} F_c^{\mu\nu_\perp} W_c n_-^\rho
  x_{\perp\sigma} \big[D_{\rm us}^\sigma,F^{\rm us}_{\rho
    \nu_\perp}\big] W_c^\dagger \right).
\end{eqnarray}  
This together with the quark Lagrangian completes the construction of
the soft-collinear effective Lagrangian including second-order power 
corrections. The Yang-Mills effective Lagrangian
is not renormalized by hard fluctuations so that the expressions given
here again hold to all orders in perturbation theory. One can now 
specify gauge fixing conditions for the collinear and ultrasoft gauge
symmetries and derive the corresponding ghost Lagrangians according to 
the standard procedure.

\paragraph{} {\bf 6.} {\bf The heavy-to-light transition current.} 
For completeness we also give the result for the representation of 
colour-singlet 
currents $\bar\psi \Gamma Q$ in the effective theory, where 
$\Gamma$ is an arbitrary Dirac matrix. These currents
appear in weak decays of heavy quarks $Q$ into light quarks. The
matching to SCET is relevant when the light quark carries large
momentum of order of the heavy quark mass. In the effective theory we
introduce a heavy quark field $h_v$ labelled by the meson velocity. 
The residual $x$ variations of this field are identical to those of
the ultrasoft light quark field, so that $h_v$ must be Taylor-expanded 
around $x_-$ in
products with collinear fields. The derivation of the
SCET current to order $\lambda^2$ and its gauge-invariant expression after
multipole-expansion for abelian gauge fields have been given in
\cite{BCDF}. With the method presented here we find for the
non-abelian case
\begin{equation}
\label{J:expansion}
\left[\bar \psi(x) \, \Gamma \, Q(x) \right]_{\rm QCD} 
=  e^{-i m v\cdot x}\, \Big\{J^{(A0)} + J^{(A1)} + J^{(A2)} + J^{(B1)}
+  J^{(B2)}\Big\}
\end{equation}
with 
\begin{eqnarray}
J^{(A0)} &=& \bar{\xi} \,\Gamma W_{c} h_v,
\\[0.2em]
J^{(A1)} &=& \bar \xi \,\Gamma W_{c} \,x_{\perp\mu} D_{\rm us}^\mu h_v  - 
 \bar \xi \,i \overleftarrow{\Slash D}_{\perp c}
     \left( i n_+ \overleftarrow{D}_{c}\right)^{-1}
     \frac{\slash n_+}{2} \Gamma W_{c} h_v,
\\
 J^{(A2)} & = &\bar \xi \,\Gamma \Wc \left( \frac12 \,n_- x\, 
n_+ D_{\rm us} h_v+ 
\frac12 \, x_{\mu \perp} x_{\nu \perp} D^\mu_{\rm us} D^\nu_{\rm us} h_v
 + \frac{i\!\!\not\!\! D_{\rm us}}{2 m} \, h_v  \right)
\\[0.1em]
  && \hspace*{-1.5cm} 
  - \, \bar \xi \,\Gamma \frac{ 1}{i n_+ D_{c}} \left[
  i n_-  D \, \Wc - W_c \,i\nm D_{\rm us} \right] h_v 
%\nonumber \\[0.1em]
%  && 
- \,\bar{\xi} \,  i\!\overleftarrow{\Slash D}_{\perp c}
     \left(i n_+ \overleftarrow{D_{c}}\right)^{-1} \frac{\slash
       n_+}{2}
     \Gamma \Wc \,x_{\perp\mu } D_{\rm us}^\mu h_v,
\nonumber\\
J^{(B1)} &=& 
  - \bar{\xi}\,\Gamma\frac{\slash{n}_-}{2m} 
  \left[i\Slash{D}_{\perp c} W_{c}\right] h_v,
\label{lambda2}
\\[0.2em]
 J^{(B2)} & = & 
- \bar \xi \, \Gamma \frac{\slash n_-}{2m} 
  \left[ i \Slash D_{\perp c} \Wc\right] 
  x_{\mu\perp} D^\mu_{\rm us}  h_v 
 - \bar \xi \, \Gamma \,  \frac{\slash n_-}{2m} \left[
  i n_-  D \, \Wc -W_c \,i\nm D_{\rm us} \right]   h_v
\nonumber\\[0.1em]
&& \hspace*{-1.5cm} - \, \bar \xi \, \Gamma 
 \frac{1}{in_+D_{c}} \left[\frac{i\Slash D_{\perp 
       c} i\Slash D_{\perp c}}{m} \Wc \right]h_v 
+ \bar{\xi}\, i\!\overleftarrow{\Slash D}_{\perp c}
     \left(i n_+ \overleftarrow{D_{c}}\right)^{-1}
     \frac{\slash n_+}{2}\Gamma 
     \frac{ \slash{n}_-}{2m}
         \left[i\Slash{D}_{\perp c}W_{c}\right]  h_v,
\label{finalJab}
\end{eqnarray}
which is identical to the result of \cite{BCDF} derived there 
for the abelian case. 
Recall that derivatives operate on ultrasoft fields before 
$x=x_-$ is set, and that derivatives with square brackets act only
within the brackets. Contrary to the effective Lagrangian
the expansion of the current may be corrected by
perturbative effects, although some relations between the various 
terms in the expansion hold due to
reparameterisation invariance. 

\paragraph{} {\bf Conclusion.} In this note we completed the
construction of soft-collinear effective theory in position 
space in terms of operators with homogeneous power counting in the 
expansion parameter $\lambda$. We first gave the exact SCET 
Lagrangian before multipole expansion and then showed how the
multipole-expansion (necessary to make the operators homogeneous in 
$\lambda$) can be constructed in a manifestly gauge invariant
form to any order in $\lambda$. The key ingredient in this
construction was to find the collinear field redefinitions, after
which the fields transform under homogeneous gauge transformations,
and to work out the $\lambda$ expansion in the intermediate fixed-line
gauge. Since the SCET Lagrangian is not
renormalized by hard interactions, this allows us in principle to 
derive the effective Lagrangian to any desired accuracy. For
heavy-to-light transition currents the result has been derived to
second order in $\lambda$ and at tree-level. 
This extends the results of \cite{BCDF}, where the
manifestly gauge-invariant form of the multipole expansion has only
been given for abelian gauge fields, and where the construction was
restricted to second order.

\vskip0.2cm
{\em Note added.}
When this work was completed, 
an article by Pirjol and Stewart appeared \cite{Pirjol:2002}, 
which addresses the gauge-invariant formulation of
power-suppressed interactions in the hybrid momentum-position space
version of SCET. The detailed comparison of our results with this
paper is complicated by two facts: first, we do not know 
the gauge transformations under which the theory is presumed to be 
invariant. (The transformations given in Table II of the
second reference of \cite{Bauer:2000yr} are obviously not appropriate,
because they are not homogeneous in $\lambda$, 
similar to (\ref{gaugetrafos})). 
Second, although the authors now also use the term ``multipole
expansion'', no multipole expansion as described around (\ref{taylor}) 
is performed in the hybrid representation.
To see how the two formulations are related, we note that in the
position representation used in the present note 
the power-suppressed terms in the effective Lagrangian do
not contain any two-point vertices (propagator corrections). 
The explicit factors of $x$ in interaction vertices 
that come from the multipole expansion act
as derivatives on propagators in momentum space and the sum of such
terms reconstructs the full QCD propagators attached to 
collinear-ultrasoft vertices order by order in 
$\lambda$. On the other hand the hybrid representation does not contain
multipole-expanded interaction vertices, but the effective Lagrangian
contains power-suppressed two-point interactions
\cite{Manohar:2002fd}. Multiple insertions
of these interactions (combined with the fact that collinear lines 
are assigned the full ultrasoft momentum absorbed at a
collinear-ultrasoft vertex in the 
hybrid representation) also lead to the reconstruction of the full
propagators. When this dictionary is applied, the expressions 
for the Lagrangian and 
heavy-to-light current given in \cite{Pirjol:2002} seem to agree 
with those of \cite{BCDF,Feldmann:2002cm} and those given here except 
for the Yang-Mills Lagrangian, where we were unable to establish 
a relation. We wish to note further that contrary
to the claim made in \cite{Pirjol:2002} the fact that no new operators
in the SCET Lagrangian are generated by radiative corrections has
already been proven in \cite{BCDF} (Section 3.4, after Eq. (63)). 
The arguments based on reparameterisation invariance given in 
\cite{Pirjol:2002} therefore provide 
an alternative derivation of this result.

\vskip0.2cm
{\em Acknowledgement.}
We thank M.~Diehl and A.~Chapovsky for useful discussions. 
%The work of M.B. is supported in part by the 
%Bundesministerium f\"ur Bildung und Forschung, Project 05 HT1PAB/2.

\end{document}